\documentclass[12pt,reqno]{amsart}
\usepackage{epsfig}
\usepackage{graphicx}
\begin{document}
\baselineskip=0.65cm 
\theoremstyle{plain}
\newtheorem{thm}{Theorem}[section]
\newtheorem{lem}{Lemma}[section]
\newtheorem{prop}{Proposition}[section]
\newtheorem{coll}{Conclusion}
\theoremstyle{remark}
\newtheorem{rem}{Remark}[section]
\title {Solitary wave solutions and their velocity selections and prohibitions for a
  general Boussinesq type fluid model}
\author{Chun-Li Chen$^1$, Sen-yue Lou$^{1,2,3}$, Yi-Shen Li$^{4}$}
\dedicatory{{$^1$Department of Physics, Shanghai Jiao Tong
University, Shanghai, 200030, P.R.China}\\{$^2$ School of
Mathematics, The University of New South Wales, Sydney, NSW 2052,
Australia}\\{$^3$Department of Physics, Ningbo University, Ningbo,
315211, P.R.China}
\\ {$^4$ Department of Mathematics, University
of Science and Technology of China, Hefei, Anhui, 230026,
P.R.China}}
\begin{abstract}
The possible solitary wave solutions for a general Boussinesq
(GBQ) type fluid model are studied analytically. After proving the
non-Painlev\'e integrability of the model, the first type of exact
explicit travelling solitary wave with a special velocity
selection is found by the truncated Painlev\'e expansion. The
general solitary waves with different travelling velocities can be
studied by casting the problems to the Newtonian quasi-particles
moving in some proper one dimensional potential fields. For some
special velocity selections, the solitary waves possess different
shapes, say, the left moving solitary waves may possess different
shapes and/or amplitudes with those of the right moving solitons.
For some other velocities, the solitary waves are completely
prohibited. There are three types of GBQ systems (GBQSs) according
to the different selections of the model parameters. For the first
type of GBQS, both the faster moving and lower moving solitary
waves allowed but the solitary waves with``middle" velocities are
prohibit. For the second type of GBQS all the slower moving
solitary waves are completely prohibit while for the third type of
GBQS only the slower moving solitary waves are allowed.

\vskip.2in

\leftline{\bf \emph{PACS.}05.45.Yv,  02.30.Jr, 02.30.Ik.}
\end{abstract}
\maketitle
\section{Introduction}
The study of the Korteweg de-Vries (KdV) equation
\begin{equation}\label{kdv}
u_t+6uu_x+u_{xxx}=0
\end{equation}
has been an interesting issue since the discovery of soliton. Its
exact solution\cite{li1} can be used to describe the overtaking
collision of soliton on a uniform layer of water, but the solution
is only physically meaningful for the unidirectional soliton. All
the left moving solitons with zero boundary conditions are
prohibited.

In Ref.\cite{wz}, three sets of model equations are derived for
modelling nonlinear and dispersive long gravity waves travelling
in two horizontal directions on shallow waters of uniform depth. A
good understanding of all solutions of these models are helpful
for coastal and civil engineers to apply the nonlinear water wave
model in a harbor and coastal design. Therefore, finding more
types of solutions of these equations are fundamental interest in
fluid dynamics. For the various different models, a point of
central interest is to examine and compare the solitary wave
solutions. For the case of plane motion, the three set of models
can be rewritten as the Boussinesq class after omitting the higher
order terms:

(A). The $\{\bar u, \ v\}$ system, -- the depth-mean velocity
basis
\begin{equation}\label{41}\begin{aligned}
       & \bar u_t+\bar u\bar u_x+v_x=\dfrac{1}{3}\bar u_{xxt} \\
       & v_t+[(1+v)\bar u]_x=0.
     \end{aligned}\end{equation}

(B). The $\{\underline{u},\ v\}$ system, -- the bottom variable
basis
     \begin{equation}\label{43}\begin{aligned}
       & \underline{u}_t+\underline{u}\underline{u}_x+v_x=\dfrac{1}{2}\underline{u}_{xxt} \\
       &
       v_t+[(1+v)\underline{u}]_x=\dfrac{1}{6}\underline{u}_{xxx}.
     \end{aligned}\end{equation}

(C). The $\{\hat u,\ v\}$ system, -- the surface variable basis
     \begin{equation}\label{45}\begin{aligned}
       & \hat u_t+\hat u\hat u_x+v_x=0 \\
       & v_t+[(1+v)\hat u]_x=-\dfrac{1}{3}\hat u_{xxx}.
     \end{aligned}\end{equation}
In the Boussinesq types of systems \eqref{41}-\eqref{45}, the
field $v$ is wave elevation and $\bar u, \ \underline{u}$ and
$\hat u$ are depth-mean, bottom and surface velocities
respectively.

To study the systems \eqref{41}-\eqref{45} uniformly, we can
extended them to the following generalized Boussinesq system
(GBQS):
\begin{equation}\label{D}\begin{aligned}
       & u_t+uu_x+v_x=c_1u_{xxt}\\
       & v_t+[(1+v)u]_x=c_2u_{xxx}.
     \end{aligned}\end{equation}

The system \eqref{45} has been shown to be integrable and has
Hamiltonian structures\cite{kaup, kup}. The exact solitary wave
and periodic wave solutions of \eqref{45} have also been obtained
by many authors\cite{ccl}. Especially, Zhang and Li\cite{zli}
presented a theory of bidirectional solitons on water by using
integrable Boussinesq surface-variable equation \eqref{45}.
However the integrability of the other two systems \eqref{41} and
\eqref{43} or generally the GBQS \eqref{D} is not known. In this
paper, we will show that the GBQS is non-Painlev\'e integrable
generally except for a special equivalent case of \eqref{45}.
Though the model is non-Painlev\'e integrable, it is still
possible and useful to give out some interesting solitary wave
solutions. In this paper, we are concentrate on to study the exact
solitary wave solutions of the GBQS \eqref{D} under the physically
significant boundary conditions
    \begin{equation}\label{boundary}
    \left.\frac{\partial^n u}{\partial x^n}\right|_{x\rightarrow \pm \infty} \rightarrow 0,
     \quad \left.\frac{\partial^n v}{\partial x^n}\right|_{x\rightarrow \pm \infty}
       \rightarrow
     0,
     \qquad n=0,\ 1,\ 2,\ \cdots.
\end{equation}
Because the integrability and the soliton solutions of the model
\eqref{45} have been known in literature, we always assume
$c_1\neq 0$ except for the special cases which will be
particularly pointed out if it is necessary.

In the section 2 of this paper, the non-Painlev\'e integrability
of the GBQS is studied by using the standard Weiss-Tabor-Carnevale
(WTC) approach\cite{WTC}. A special explicit solitary wave
solution with a specific velocity selection is given by the
truncated Painlev\'e expansion. In Sec. 3, after reflecting the
problem to find the possible solitary waves to the possible
motions of the Newtonian type quasi-particles moving in some
proper potential fields, the velocity prohibition and selection
phenomena for the solitary waves of the GBQS are discussed. For
the general solitary wave with allowed velocities, an implicit
form of the exact solitary wave solutions is given. Some special
solitary wave solutions are plotted also according to the implicit
expression. The last section is a short summary and discussions.

\section{Non-Painlev\'e integrability  of the GBQS and its explicit exact solitary
waves with the first type of special velocity selection}

{\bf 2.1 Non-Painlev\'e integrability of the GBQS with $c_1\neq
0$}

In the modern soliton theory, the study of the
Painlev\'e property\cite{WTC} plays a very important role because
it can be used not only to isolated out (Painlev\'e) integrable
models\cite{Pain} but also to find many other integrable
properties such as the B\"acklund transformations, Lax pair,
Schwarzian form etc\cite{WTC,ccl}. Furthermore, even if the model
is non-Painlev\'e integrable, the method can still be used to find
some useful things such as the special exact explicit
solutions\cite{Lou}. Because it has been known that the model
\eqref{45} (i.e., \eqref{D} with $c_1=0$) is integrable and the
general real physical model may require $c_1\neq 0$, say,
\eqref{41} and \eqref{43}, we only check the Painlev\'e property of
the model with $c\neq0$ in this subsection and find a special solitary wave
solution from the next subsection.

The Painlev\'e property of a partial differential equation system
is defined as all the solutions of the system are free of the
essential and branch movable singularities around an arbitrary
(both characteristic and non-characteristic) singular
manifold\cite{chara}.

According to the above definition of the Painlev\'e property, the
general solutions of the GBQS should have the following expansion
around the arbitrary singular manifold $\phi\equiv \phi(x,y,t)=0$
\begin{equation}\label{uvw}
u=\sum^\infty_{j=0}u_j\phi^{j+\alpha}, \quad
v=\sum^\infty_{j=0}v_j\phi^{j+\beta}
\end{equation}
where $\{\alpha, \ \beta\}$ should be negative finite integers.
Furthermore, if the model possesses the Painlev\'e property, there
should be at least one primary branch (one possible selection of
$\{\alpha,\ \beta,\ u_0,\ v_0\}$) such that enough arbitrary
functions (six for the GBQS \eqref{D}, $\phi$ and five of $\{u_j,\ v_j\}$)
 can be included in the Painlev\'e expansions.

By using the standard leading order analysis, i.e. substituting
$u=u_0\phi^{\alpha},\ v=v_0\phi^{\beta}$ into the GBQS and
balancing the leading terms, we can find that there is only one
possible selection
\begin{equation}\label{alpha}
\alpha=-2, \quad \beta=-2,\qquad u_{0} =12c_1\phi_x\phi_t, \quad
v_0 = 6c_2\phi_x^2
\end{equation}
for $c_2\neq 0$. For $c_2=0$, there is no possible negative
integer selection for $\alpha$ and $\beta$.

Substituting the above expansion\eqref{uvw} with \eqref{alpha} to
Eq.\eqref{D} and vanishing all the coefficients of the powers
$\phi^k$, we can find the following recursion relation for
determining the expansion functions $u_j$ and $v_j$
\begin{eqnarray}
&&(j-4)(j-6)(1+j)u_j=F_{1j},\label{recursion1}\\
&&\phi_x(j-4)c_2j(j-5)u_j-12\phi_t(j-4)c_1v_j=F_{2j},\label{recursion2}
\end{eqnarray}
where $F_{1j}$ and $F_{2j}$ are all only functions of $u_0,\ u_1,\
...,\ u_{j-1},\ v_0,\ v_1,\ ...,\ v_{j-1}$ and the derivatives of $\phi$. From the
recursion relations \eqref{recursion1} and \eqref{recursion2}, we
know that all the functions $u_j$ and $v_j$ are fixed by the
recursion relations except for those related to the resonance
points determined by vanishing the coefficient determinant of
\eqref{recursion1} and \eqref{recursion2}:
 \begin{equation}\label{det}
 \left|
\begin{array}{cc}
(j-4)(j-6)(1+j) & 0\\
\phi_x(j-4)c_2j(j-5) & -12\phi_t(j-4)c_1
\end{array}
 \right|
 = -12\phi_tc_1(j-4)^2(j+1)(j-6)=0.
 \end{equation}
From the expression \eqref{det} and by checking the resonance
conditions at the resonant points $j=4,\ 4$ and $6$, we know that
there are two facts which destroy the Painlev\'e integrability of
the GBQS with $c_1\neq 0$. The first one is that there is no
primary branch at all because the both equations of the QBQS
\eqref{D} are three order, so the primary branch should have six
resonant points such that six arbitrary functions can be entered
into the general Painlev\'e expansion solution. The second one is
the resonance conditions at $j=4$ and $j=6$ are not satisfied. The
lack of the primary branch of the model means some kinds of the
logarithmic branch and/or essential singularities around the
singular manifold $\phi=0$ may appear for the GBQS \eqref{D} with
$c_1\neq0$. Actually, substituting $u=u_0\phi^{\alpha},\
v=v_0\phi^{\beta}$ into the GBQS \eqref{D}, we can find that there
exists a possible non-completely negative selection for the
constants $\alpha$ and $\beta$:
\begin{equation}\label{alpha1}
\alpha=-2, \quad \beta=0,
\end{equation}
for both $c_2\neq 0$ and $c_2=0$ cases! The allowance of the
selection $\beta=0$ means that the logarithmic term $\ln \phi$
should be included in the expansion of the function $u$ to balance
the leading order terms of the first equation of \eqref{D}.

According to the above discussions, we can conclude that the GBQS
with $c_1\neq 0$ has no Painlev\'e property. In other words, the
model is non-Painlev\'e integrable. Though the model is
non-Painlev\'e integrable, it is still possible to find some
useful information from the Painlev\'e analysis. In the next
subsections we are specially interested to find the possible
solitary wave solution from the truncated Painlev\'e expansion.

{\bf 2.2 Truncated Painlev\'e expansion and explicit exact
solitary waves with special velocity selections}

According to the discussion of the last subsection, we know that
the standard truncated Painlev\'e expansion has the form
\begin{equation}\label{trun}
u=u_0\phi^{-2}+u_1\phi^{-1}+u_2, \quad
    v=v_0\phi^{-2}+v_1\phi^{-1}+v_2,
\end{equation}
where $\{u_2,\ v_2\}$ is a seed solution of the GBQS \eqref{D} and
usually is taken as constant solution to get the single travelling
solitary wave solution of the model.

As usual, substituting the truncated expansion \eqref{trun} with
constants $u_2$ and $v_2$ into the GBQS \eqref{D} and vanishing
the coefficients for different orders of $\phi$, we have
\begin{eqnarray}
&&u_{0} =12c_1\phi_x\phi_t, \quad v_0 = 6c_2\phi_x^2,\label{uv0}\\
&&u_{1} =-\frac{12}5\left(\frac{\phi_x\phi_{tt}}{\phi_t}
+\frac{\phi_t\phi_{xx}}{\phi_x}+3\phi_{xt}\right),\label{u1}\\
&&v_{1} =\frac{2}5\left(2\frac{\phi_x^2\phi_{tt}}{\phi_t^2}
-\frac{4\phi_x\phi_{xt}}{\phi_t}-13\phi_{xx}\right),\label{v1}
\end{eqnarray}
for the functions $u_0,\ v_0,\ u_1,\ v_1$ while the function
$\phi$ should be determined from the following over-determined
system
\begin{eqnarray}
&&u_2u_{1x}+v_{1x}+u_{1t}-c_1u_{1xxt}=0,\label{sym1}\\
&&v_2u_{1x}+v_{1t}+u_2v_x+u_x-c_2u_{1xxx}=0,\label{sym2}
\end{eqnarray}
\begin{eqnarray}
&&u_{0t}-u_1\phi_t+c_1(2\phi_xu_{1xt}+u_{1t}\phi_{xx}+(u_{1}\phi_{t})_{xx}-u_{0xxt})
\nonumber\\
&& \quad
+u_1u_{1x}-u_2u_1\phi_x+u_2u_{0x}-v_1\phi_x+v_{0x}=0,\label{f21}
\end{eqnarray}
\begin{eqnarray}
&&(u_0u_1)_x+2c_1[\phi_{xx}u_{0t}+(\phi_tu_{0})_{xx}-\phi_{xx}\phi_tu_1-u_{1t}\phi_x^2]
\nonumber\\
&&\quad
-2u_0\phi_t+[24c_1(u_{0t}-u_1\phi_{t})_x-2v_0-u_1^2-2u_2u_0)]\phi_x=0,\label{f31}
\end{eqnarray}
\begin{eqnarray}
&&c_2[3(\phi_xu_{1x})_x+u_1\phi_{xxx}-u_{0xxx}]-(u_1+u_1v_2+v_1u_2)\phi_x\nonumber\\
&& \quad
(1+v_2)u_{0x}+v_{0t}+v_{0x}u_2+(v_1u_1)_x-v_1\phi_t=0,\label{f22}
\end{eqnarray}
and
\begin{eqnarray} &&
(u_0v_1+v_0u_1)_x-2v_0\phi_t-2(v_2u_0+v_0u_2+v_1u_1+u_0)\phi_x\nonumber\\
&& \quad
+2[u_0\phi_{xxx}-3\phi_x(u_1\phi_x)_x+3(\phi_xu_0)_x]c_2=0.
\label{f32}
\end{eqnarray}
To find all the possible exact solutions of the over-determined
system is quite difficult. However, it is a quite easy work to
find the travelling solitary wave from the system
\eqref{uv0}--\eqref{f32}. For the travelling wave solution
$\phi=\phi(x-ct)\equiv \phi(\tau) $, we have
\begin{equation}\label{travel}
\phi_x=\phi_\tau,\qquad \phi_t=-c\phi_{\tau},
\end{equation}
with $c$ being an arbitrary velocity parameter.

Under the travelling wave solution condition \eqref{travel},
\eqref{uv0}--\eqref{v1} are simplified to
\begin{eqnarray}\label{uv01}
u_{0} =-12c_1c\phi_\tau^2, \quad v_0 = 6c_2\phi_\tau^2,\ u_{1}
=12cc_1\phi_{\tau\tau},\qquad v_{1}=-6c_2\phi_{\tau\tau}.
\end{eqnarray}
Substituting \eqref{uv01} into \eqref{sym1} and \eqref{sym2}
yields
\begin{eqnarray}
&&2c^2c_1^2\phi_{\tau\tau\tau\tau\tau}+(2u_2cc_1-2c^2c_1-c_2)
\phi_{\tau\tau\tau}=0,\label{phi1}\\
&&2cc_1c_2\phi_{\tau\tau\tau\tau\tau}+[2cc_1(1+v_2)+c_2(c-u_2)]\phi_{\tau\tau\tau}=0.
\label{phi2}
\end{eqnarray}

The equation system \eqref{phi1} and \eqref{phi2} is consistent only for
\begin{eqnarray} \label{uv2}
v_2=\frac{c_2^2+2c^2c_1^2(c_2k^2-2)}{4c^2c_1^2},\
u_2=\frac{c_2^2+2c^2c_1^2(c_2k^2-2)}{2cc_1c-2+2c_1c^2(1-c_1k^2)}\
\end{eqnarray}
where the constant $k$ is introduced for convenience later. Using the constant
relation \eqref{uv2},
the general solution of
\eqref{phi1} (and \eqref{phi2}) reads
\begin{eqnarray} \label{rphi}
\phi=b_0+b_1\tau+b_2\tau^2+a_1\exp(k\tau)+a_2\exp(-k\tau)
\end{eqnarray}
with $b_0,\ b_1,\ b_2,\ a_0$ and $a_1$ being arbitrary constants.
Substituting \eqref{uv01} with \eqref{rphi} into \eqref{f21}, we find that the
constants appeared in
\eqref{rphi} have to be fixed by
\begin{eqnarray} \label{ab}
b_1=b_2=a_1a_2=0.
\end{eqnarray}
Furthermore, without loss
of generality, the constant $a_2$ can be taken as zero because $k$ is arbitrary
(can be taken as both positive and negative).
After finishing some simple direct calculation, one can find that \eqref{rphi} with
  \eqref{ab}
solves all other remained equations \eqref{f31}--\eqref{f32}.

Substituting \eqref{uv01}, \eqref{uv2}, with \eqref{rphi}, \eqref{ab} and $a_2=0$ into
\eqref{trun}, we get a special solitary wave
 \begin{eqnarray}
 u& =&   \frac{c_2^2+2c^2c_1^2(c_2k^2-2)}{2cc_1c-2+2c_1c^2(1-c_1k^2)}
 +3c c_1k^2 {\rm sech}^2\left[\frac k2(\tau-\tau_0)\right],\label{41u}\\
   v &=&\frac{c_2^2+2c^2c_1^2(c_2k^2-2)}{4c^2c_1^2}-\frac32 c_2k^2 {\rm
   sech}^2
   \left[\frac k2(\tau-\tau_0)\right].\label{41v}
\end{eqnarray}
where $\tau_0=\ln(b_0)-\ln(a_1)$. In general, the boundary
conditions expressed in \eqref{boundary} can not be satisfied
 for the solitary wave solution expressed by
\eqref{41u} and \eqref{41v}. In order to
satisfy the boundary conditions \eqref{boundary} for \eqref{41u} and \eqref{41v}, we have to fix the
constants $k$ and $c$ as
\begin{eqnarray} \label{kc}
k^2=\frac{c_0^2+2}{2c_2}, \ c^2=\frac{c_0^4}{(2-c_0^2)},   \ c_0^2\equiv \frac{c_2}{c_1},
\end{eqnarray}
and then the solitary wave solution becomes
 \begin{eqnarray}
 u &=&\pm \dfrac{3(c_0^2+2)}{2\sqrt{2-c_0^2}}{\rm sech}^2\left[\dfrac{\sqrt{2c_1(c_0^2+2)}}
 {4c_1c_0}\left(x\mp
    \dfrac{c_0^2}{\sqrt{2-c_0^2}}t-\tau_0\right)\right],\label{41su}\\
   v &=&-\dfrac{3}{4}(c_0^2+2){\rm sech}^2\left[\dfrac{\sqrt{2c_1(c_0^2+2)}}{4c_1c_0}
   \left(x\mp
    \dfrac{c_0^2}{\sqrt{2-c_0^2}}t-\tau_0\right)\right]. \label{41sv}
\end{eqnarray}

From the above discussions, the zero
boundary travelling solitary wave solution \eqref{41su}  (and  \eqref{41sv})
obtained via the truncated Painlev\'e expansion is valid only for the GBQS
\eqref{D} with the condition $c_0^2<2\ ({\rm i.e.,} \ c_2<2c_1) $
and $c_2\neq0$ for $c_1\neq0$. The solitary wave solution
\eqref{41su} (and  \eqref{41sv}) possesses only two special isolated velocities
$\pm \frac{c_0^2}{\sqrt{2-c_0^2}}$. Is there any other travelling solitary
 wave with other velocities for the GBQS \eqref{D}? The answer is positive because
 in the truncated Painlev\'e
 expansion approach we require all the coefficients of the different powers of $\phi$ being
   zero.
 This strong condition leads to the loss of the generality. In the
 next section, we discuss this problem generally by casting
 the problem to the possible motions of a Newtonian classical
 quasi-particle in some possible potentials.

\section{Implicit travelling solitary wave solutions and special
velocity selections and prohibitions}

From the last section, we know that by means of the truncated
Painlev\'e expansion approach, we can only obtain a special exact
solitary wave solution with the boundary condition
\eqref{boundary}. Actually, the travelling solitary waves with
different velocities do exist for the GBQS \eqref{D}. Though we
can not explicitly write down the exact solitary wave solutions
with the boundary condition \eqref{boundary} for the different
velocities, we can find an uniform and implicit formula for the
solitary wave solutions except for some isolated special critical
velocities.

For a travelling wave solution, $u=u(x-ct)\equiv u(\tau),\
v=v(\tau)$, the motion equation system \eqref{D} becomes an
ordinary differential equation system
\begin{equation}\label{D1}
\begin{aligned}
       & -c u+\frac{1}2 u^2+ v=-c_1   c u_{\tau\tau}\\
       & -c v+ (1+v)u=c_2  u_{\tau\tau},
     \end{aligned}
\end{equation}
where the both equations have been integrated once with respect to
$\tau$ and the integration constants have been fixed as zero because
of the boundary conditions given in \eqref{boundary}. From the
first equation of \eqref{D1}, we know that the travelling wave of
the field $v$ can be simply expressed by
 \begin{equation}\label{rv}
     v=-\dfrac{1}{2}(-2c u+u^2+2c_1 c u_{\tau \tau}).
   \end{equation}
Substituting it into the second equation of \eqref{D1} we have
\begin{eqnarray}\label{ruxx}
u_{\tau\tau}=\frac{2(1-c^2)u- u^3+3c u^2}{2c_1 (c_0^2 -c^2+c u)}.
\end{eqnarray}
The first integration of \eqref{ruxx} reads
\begin{eqnarray}\label{ru}
u_\tau^2&=&\frac{1}{ c_1}\left(\frac{c_0^2}{c^2}-1\right)
\left(\frac{c_0^4 }{c^2}+c_0^2-2\right)\ln
\frac{cu+ c_0^2-c^2}{ c_0^2-c^2}\nonumber \\
&&
-\frac{u^3}{3c_1c}+\frac{u^2}{2c_1}\left(2+\frac{c_0^2}{c^2}\right)
-\frac{u}{c_1c}\left(\frac{c_0^4 }{c^2}+c_0^2-2\right)\nonumber\\
&\equiv &-2 V(u)
\end{eqnarray}
for $c\neq0$ and
\begin{equation}\label{ru0}
u_\tau^2=-\frac{u^4}{4c_2}+\frac{u^2}{{c_2}}\equiv -2 V0(u),\
\tau=x,\ (c_2\neq0)
\end{equation}
for $c=0$. To get the relations \eqref{ru} and \eqref{ru0}, the
integration constants have been fixed appropriately such that the
boundary conditions \eqref{boundary} may be satisfied.

Now, from the expressions \eqref{ruxx}--\eqref{ru0}, it is known
that to find the possible travelling solitary wave solutions of
the GBQS \eqref{D} with boundary condition \eqref{boundary} is
equivalent to find the possible special motions of a classical
quasi-particle moving in the potential fields $V(u)$ and/or
$V0(u)$ related to the \em maximum \rm point at $u=0$ with the
``space" variable $u$ and the ``time" variable
$\tau$\cite{instanton}.

If the existence problem of the solitary wave is solved, except
for some special critical cases (see later), the travelling
solitary waves with the boundary condition \eqref{boundary} can be
uniformly expressed implicitly by
\begin{equation}\label{imp}
\tau-\tau_0=\pm \int^u\frac{{\rm d}u}{\sqrt{-2V(u)}},
\end{equation}
where $V(u)$ is defined in \eqref{ru} and the integration constant
$\tau_0$ is related to the location of the solitary wave.

To see the possible solitary wave solutions qualitatively, the
possible motions of the quasi-particle in the potentials $V(u)$
and/or $V0(u)$, it is convenient to study the structures of the
potentials $V(u)$ and $V0(u)$ at the same time.

To study the structures of the potentials $V(u)$, we firstly
isolate out some special and/or critical cases.

The first special case is same as that in \eqref{kc}. When the
velocity $c$ of the solitary wave is fixed by \eqref{kc}, the
logarithmic term of \eqref{ru} vanishes and the function $u$ can
be explicitly integrated out. The result is reasonably same as
that obtained in the last section via the truncated Painlev\'e
analysis. The potential structure at this special case is plotted
in Fig. 1a for the special model \eqref{43}, i. e. $c_1=\frac12,\
c_2=\frac16$ and then $c_0=\sqrt{\frac13}$ and
$c=\sqrt{\frac1{15}}$. The corresponding soliton solution for the
field $u$ is plotted in Fig.1b.

\input epsf
\begin {figure}
\centering \epsfxsize=7cm\epsfysize=5cm\epsfbox{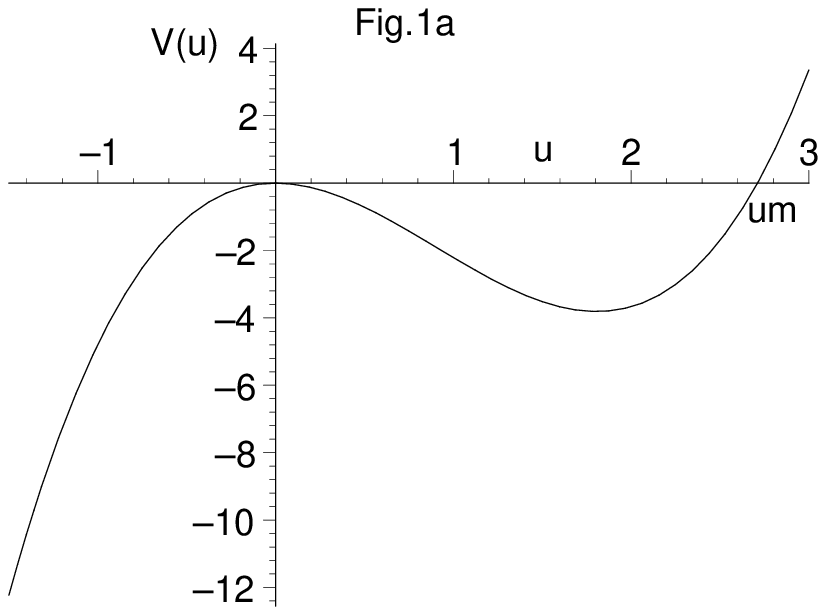}
\epsfxsize=7cm\epsfysize=5cm\epsfbox{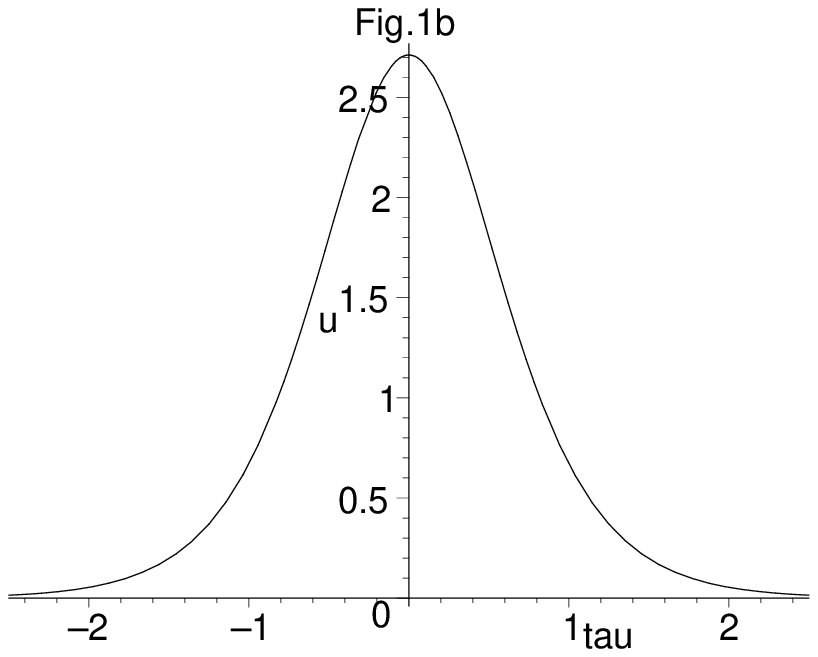} \caption{(a). The
potential structures for the special velocity selection
$c=\frac{c_0^2}{\sqrt{2-c_0^2}}$ with $c_1=\frac12$ and $
c_0=\frac{\sqrt{3}}3$ ($c_2=\frac16$). (b) The bell shape solitary
wave solution related to (a). All the figures of this paper have no unit because the model
is dimensionless. $tau \equiv \tau$ in all figures also.}
\end{figure}

The second special critical case is that if the velocity is
determined by
\begin{equation}\label{ro1}
c=\pm c_0,
\end{equation}
then the logarithmic term of \eqref{ru} is also vanished. However,
in this case there is no solitary wave solution with the boundary
condition \eqref{boundary} except for $c_0=1$, i.e., $c_1=c_2$. In
other words, the solitary wave solutions with the velocities $\pm
c_0$ are prohibited for the GBQS \eqref{D} with $c_2\neq c_1,\
c_1\neq 0$. This velocity prohibition property can be observed
more clearly from \eqref{ruxx}. After substituting \eqref{ro1}
into \eqref{ruxx}, one can directly see that it is impossible to
satisfy the boundary condition \eqref{boundary} except for
$c_0=1$. While if $c_0=1$, \eqref{ro1} is same as \eqref{kc} and
then the related solitary wave solution is also given by
\eqref{41su}. The related potential structure of the
potential $V(u)$ with $c_1=\frac12,\ c_2=\frac16$ and then
$c=c_0=\sqrt{\frac13}$ is plotted in Fig. 2. From Fig. 2, we can
see also that when $c=c_0$, $u=0$ is not an extremum of the
potential while a solitary wave is corresponding to a possible
special motion of the quasi-particle related to a maximum of the
potential.

\input epsf
\begin {figure}
\epsfxsize=7cm\epsfysize=5cm\epsfbox{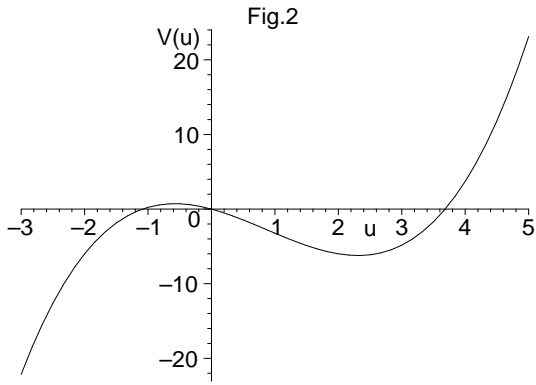} \caption{The
potential structures for the critical velocity $c=c_0$ with
$c_1=\frac12$ and $ c_0=\frac{\sqrt{3}}3$.}
\end{figure}

The third special critical case is related to the static solitary
wave solution. For the static solitary wave solution of the GBQS,
two subcases, $c_2>0$ and $c_2\leq 0$, should be clarified. For
the GBQS \eqref{D} with $c=0,\ c_2\neq0$, \eqref{ru} is not valid
and has to be changed as \eqref{ru0}. The corresponding structure
for the potential $V0(u)$ is plotted in Fig. 3a for
$c_2=\frac16>0$ and the corresponding solitary wave solution is
plotted in Fig. 3b.

From \eqref{ru0} and \eqref{rv}, it is straightforward to get that
the static solitary wave of the GBQS with $c_2>0$ has the form
\begin{equation}\label{u0}
u=\pm 2{\rm sech}\frac{x}{\sqrt{c_2}},\ v=-{\rm
sech}^2\frac{x}{\sqrt{c_2}}.
\end{equation}

\input epsf
\begin {figure}
\centering \epsfxsize=7cm\epsfysize=5cm\epsfbox{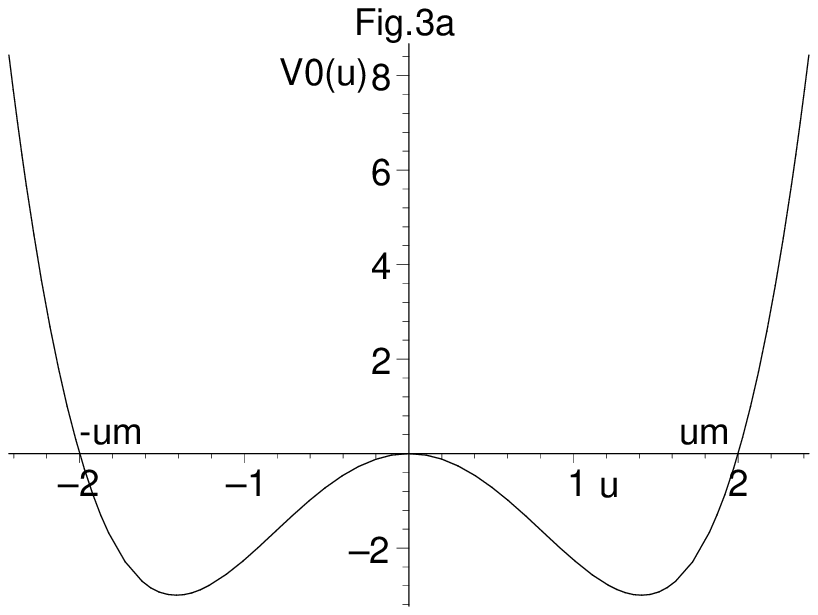}
\epsfxsize=7cm\epsfysize=5cm\epsfbox{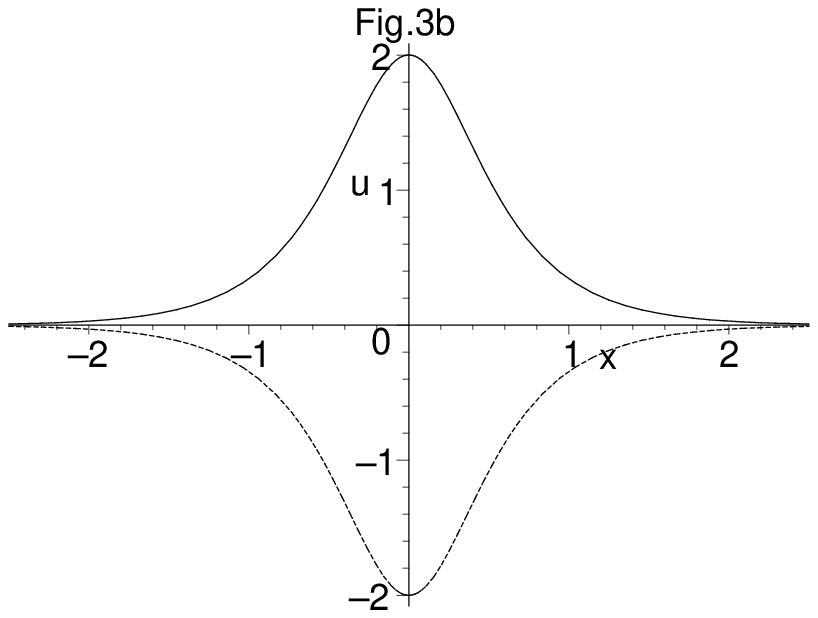} \caption{(a) The
potential structures for the static velocity $c=0$ with
$c_2=\frac16>0$;. (b) The related static bell shape solitary wave
(up solid line) and ring shape wave (the lower dashed line).}
\end{figure}

Fig. 4 shows the structure for the potential $V0(u)$ with
$c_2=-\frac13<0$ which includes the integrable case \eqref{45}.
From \eqref{ru0} and Fig. 4, it is also easy to see that whence
$c_2<0$, $u=0$ is a minimum of the potential $V0(u)$. So there is
no static solitary wave of the GBQS with $c_2<0$ and zero boundary
conditions because a solitary wave corresponding to the motion of
the classical quasi-particle in the potential field related to a
\em maximum. \rm  Furthermore, from \eqref{D1} with $c=0$ and
$c_2=0$, we can also know that there is no static solitary wave at
all. In summary, the static solitary wave with zero boundary
condition is prohibited for the GBQS with $c_2\leq 0$. Especially,
there is no static solitary wave with zero boundary conditions for
the systems \eqref{41} and \eqref{45} even if \eqref{45} is
integrable.

\input epsf
\begin {figure}
\centering \epsfxsize=7cm\epsfysize=5cm\epsfbox{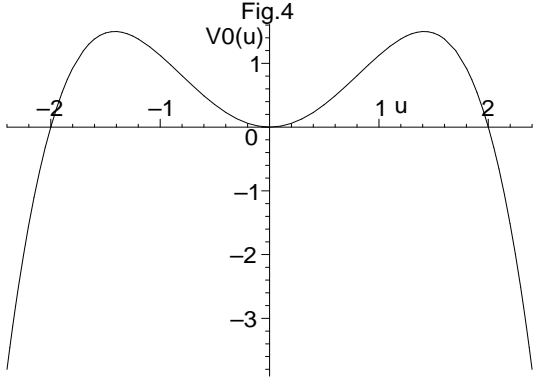}
\caption{The potential structures for the static velocity $c=0$
with $c_2=-\frac13<0$.}
\end{figure}

The fourth type of velocity selection is related to
\begin{equation}\label{ro2}
c=\pm 1,\qquad c_0\neq 1.
\end{equation}
Usually, the solitary waves are decay exponentially apart from the
solitary wave center. However, for the GBQS with $c_2\neq c_1$, if
the solitary wave possesses the velocities $\pm 1$ then it decays
only algebraically. This velocity selection property can be seen
from the second order differentiation with respect to $u$ of the
potential $V(u)$ at the point $u=0$. The second order
differentiation of $V(u)$ for $c^2=1$ reads
\begin{equation}\label{inflexion}
\frac{{\rm d}^2V(u)}{{\rm
d}u^2}=-\frac{u[6(c-u-cc_0^2)+2cu^2+3uc_0^2]}{2c_1(1-cu-c_0^2)^2},\
(c^2=1).
\end{equation}
It is known that the boundary values ($u=0$ in our case) of
exponentially decayed solitary wave solutions are linked with the
maxima of the potential $V(u)$\cite{instanton}. However, from
\eqref{inflexion} we know that when the velocity $c=\pm 1$, $u=0$
is only an inflexion point of the potential $V(u)$. It is also
known that an inflexion point may be related to a rational (or
algebraic) solitary wave solution\cite{sG}. This type of the
algebraic solitary wave is similar to some what of the solitary
wave at the critical point where the phase transition occur for
some types of quantum fields and condense matter systems\cite{sG}.

To see the algebraic decay property of the solitary waves at this
critical case, we take $c=1,\ c_1=\frac12$ and $c_0=50$. Under
this parameter selection, the related potential becomes
\begin{eqnarray}
V(u)&=&15624992502\ln\left(\frac{u}{2499}+1\right)-\frac13u^3+1251u^2-6252498u\equiv
V1(u) \label{approV1} \\
 &\approx& \frac1{2499}u^3-\frac{139}{1387778}u^4\equiv V2(u). \label{approV2}
\end{eqnarray}
The corresponding structure for the exact potential $V1(u)$ and
the approximate potential $V2(u)$ are plotted at the same figure,
Fig. 5a. From Fig. 5a, it is seen that two lines are almost
coincide with each other at the plotted region which related to
the quasi-particle move in. Actually, the approximation
\eqref{approV2} is quite well up to $u\sim \pm 1000$ while the
solitary wave is located only at the region $u=0\sim 4$. So to see
the solitary wave structure in this special case we can safely use
the approximate potential $V2(u)$ to calculate the related
solitary wave. The result reads
\begin{equation}\label{appro}
u_{c=1, c_1=0.5,\ c_2=100} \approx \frac{4998}{\tau^2+1251}.
\end{equation}
Fig. 5b. shows the structure of the algebraic solitary wave
expressed by \eqref{appro}.

\input epsf
\begin {figure}
\centering \epsfxsize=7cm\epsfysize=5cm\epsfbox{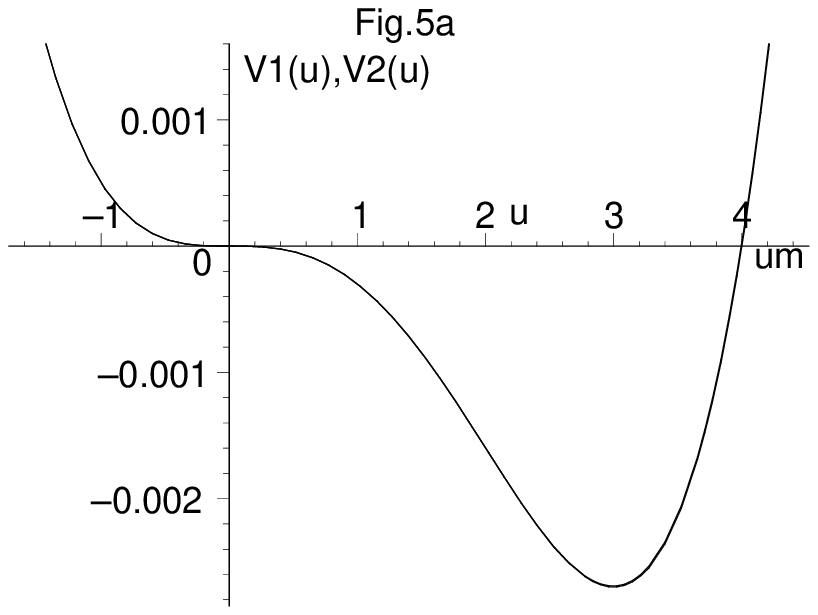}
\epsfxsize=7cm\epsfysize=5cm\epsfbox{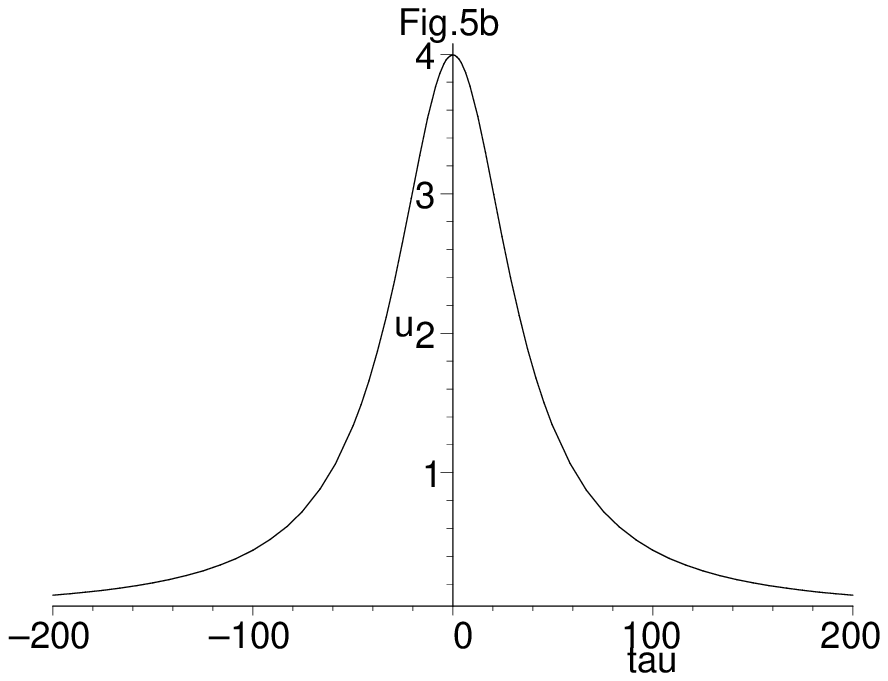} \caption{(a). The
potential structures for the potentials $V1(u)$ and $V2(u)$ at the critical
velocity $c=1$ with $c_1=\frac12$ and $c_0=50$. (b). The corresponding approximately algebraic decayed solitary
wave. }
\end{figure}

From the analysis of the existence problem of the solitary waves
at the critical velocities, we know that for different velocities
the solitary waves may possesses different shapes. This type
phenomena occurs also in integrable cases. For instance, for the
bidirectional Kaup-Kupershmidt equation, its right moving solitons
possess single-humped shape while its left moving solitons possess
double-humped shape\cite{bkk}. In integrable case, the velocity
prohibition phenomena are also common. For instance, for the
system \eqref{45} the static solitary wave with zero boundary
condition is prohibited and for the KdV equation \eqref{kdv}, all
the left moving and static soliton solutions with zero boundary
conditions are prohibited.

Now the important problem is in addition to the selections and
prohibitions at the critical points whether there exist further
prohibitions for some other velocity regions. To solve this
problem, we restrict ourselves for $c_1>0$ (because most of the
real physical systems, say, \eqref{41} and \eqref{43}, possess
this property) and $c>0$ (because the symmetry property $\{c,\
u\}\leftrightarrow \{-c,\ -u\}$ of \eqref{ruxx}).

To find the existence conditions of the solitary waves  of the
GBQS \eqref{D} with zero boundary conditions \eqref{boundary} for
noncritical cases, we can use the maximum condition, $\left.
\frac{{\rm d^2} V(u) }{{\rm d} u^2}\right|_{u=0}<0$ and the real
condition of the potential $V(u)$.

The maximum condition of the potential $V(u)$ at $u=0$ reads
\begin{equation}\label{vuu}
\left. \frac{{\rm d^2}V(u)}{{\rm d}
u^2}\right|_{u=0}=\frac{1-c^2}{c_1(c^2-c_0^2)}<0,
\end{equation}
and the real condition of $V(u)$, reads
\begin{equation}\label{real}
\frac{cu}{c_0^2-c^2}+1>0.
\end{equation}
After finishing the  detail analysis with help of the conditions
\eqref{vuu} and \eqref{real}, we find the following six different
structures for the potential $V(u)$ for $c_1>0$:

\em Case 1.\rm
\begin{equation}\label{case1}
 0<c^2<c_0^2<1.
\end{equation}
When $c^2$, the square of the velocity parameter, is located in
the range $(0,\ c_0)$ with $c_0<1$, the related potential $V(u)$
possesses the structure as shown in Fig. 6a for $c>0$. From Fig.
6a we know that there exist a special motion for a classical
quasi-particle moving in this potential related to the maximum at
$u=0$: At the beginning ($\tau=-\infty$), the quasi-particle is
located at the peak center $u=0$, as time ``$\tau$" increases, the
quasi-particle ``roll" down the hill up to the $um\equiv u_{am}$
point and then return back to the original point $u=0$ at ``time"
$\tau=+\infty$. This type of special motion of the quasi-particle
is just related to the solitary wave solution of the GBQS
\eqref{D} with \eqref{case1}. From Fig. 6a we know also that the
quasi-particle can roll only to right (the positive $u$ direction)
that means the right moving solitary wave ($c>0$) possesses bell
shape for the field $u$. By using the invariant transformation of
\eqref{ruxx}, $\{u,\ c\}\rightarrow \{-u,\ -c\}$, we know that the
left moving solitary wave ($c<0$) possesses ring shape. The
corresponding exact solitary wave with the same parameters as
shown in Fig. 6a is plotted in Fig. 6b.

\input epsf
\begin {figure}
\epsfxsize=7cm\epsfysize=5cm\epsfbox{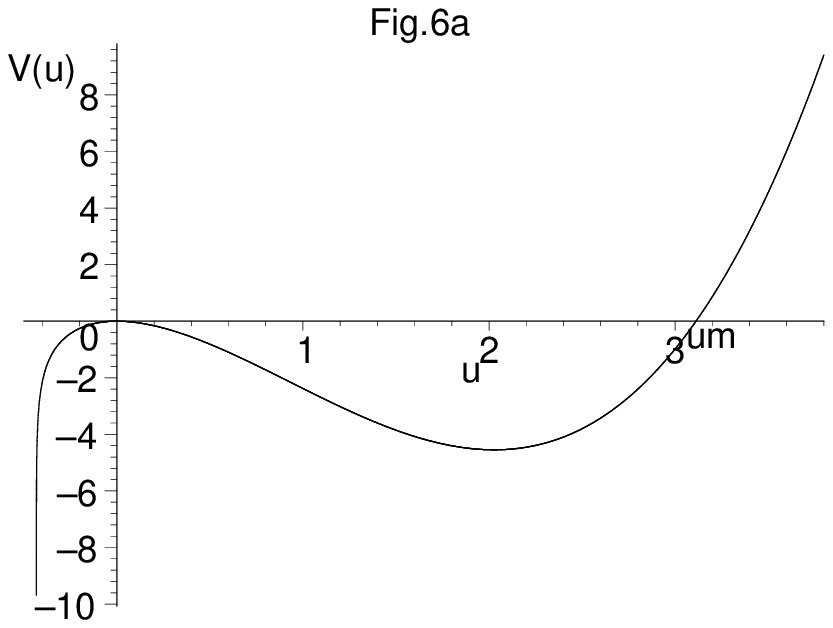}
\epsfxsize=7cm\epsfysize=5cm\epsfbox{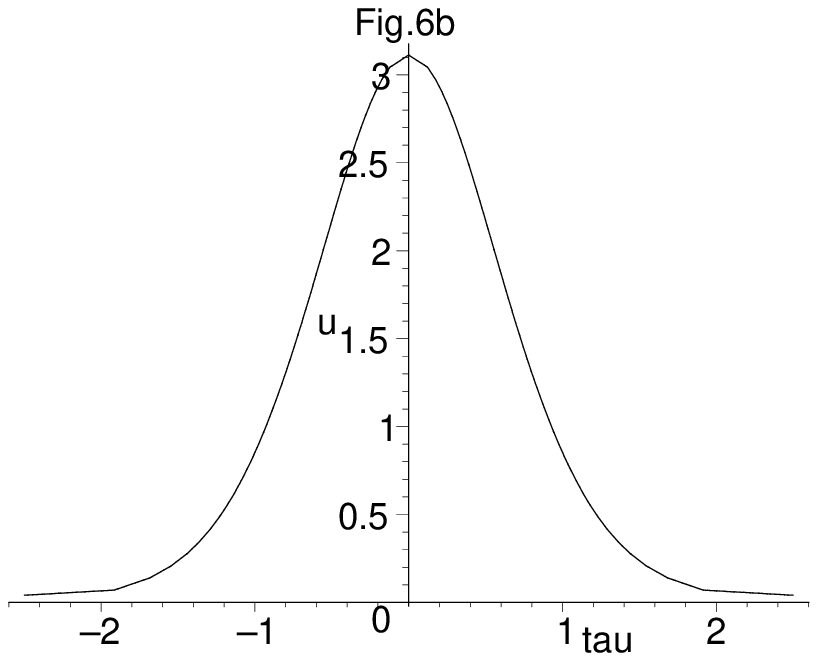}
 \caption{(a) The
typical structure for the potential of the quasi-particle located
in with $c_1>0$ and $c\geq0$, $0<c^2<c_0^2<1$ and $\{
c_0=\frac{1}{\sqrt{3}},\ c=0.4,\ c_1=\frac12\}$. (b) The
corresponding solitary wave solution related to (a).}
\end{figure}

\em Case 2.
\begin{equation}\label{case2}
c_0^2<c^2<1.
\end{equation}
\rm In this case, the related potential $V(u)$ possesses the
structure as shown in Fig. 7 for $c>0$. From Fig. 7 we know that
there is no solitary wave solution with zero boundary condition in
this case because $u=0$ is only a minimum of the potential. If the
quasi-particle stay at $u=0$ at the beginning then it can only be
stayed there forever. In other words, the solitary waves of the
GBQS \eqref{D} with the velocities $c_0^2<c^2<1$ and zero boundary
condition \eqref{boundary} are totally prohibited.

\input epsf
\begin {figure}
\epsfxsize=7cm\epsfysize=5cm\epsfbox{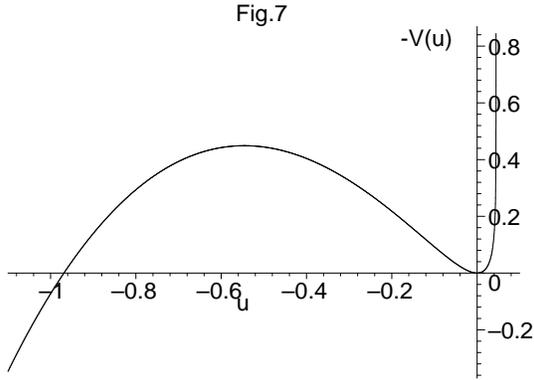}\caption{The
typical structures for the potential with $c_1>0$ and $c\geq0$
$c_0^2<c^2<1$  and $\{c_0=\frac{1}{\sqrt{3}},\ c=0.6,\
c_1=\frac12\}$.}
\end{figure}

\em Case 3. \rm
\begin{equation}\label{case3}
c_0^2<1,\ c^2>1.
\end{equation}
The structure of the potential $V(u)$ for $c^2>1$ with $c_0^2<1$
possesses the form as shown in Fig. 8a. Similar to the case 1,
there exist the right moving bell shape solitary waves and the
left moving ring shape solitary waves for the field $u$. The right
moving bell shape solitary wave solution related to Fig. 8a is
plotted in Fig. 8b.

\input epsf
\begin {figure}
\epsfxsize=7cm\epsfysize=5cm\epsfbox{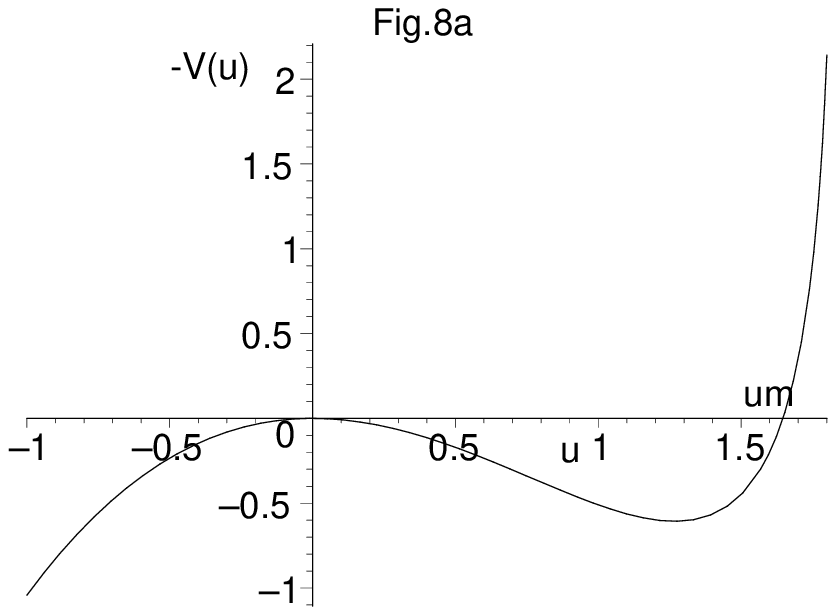}
\epsfxsize=7cm\epsfysize=5cm\epsfbox{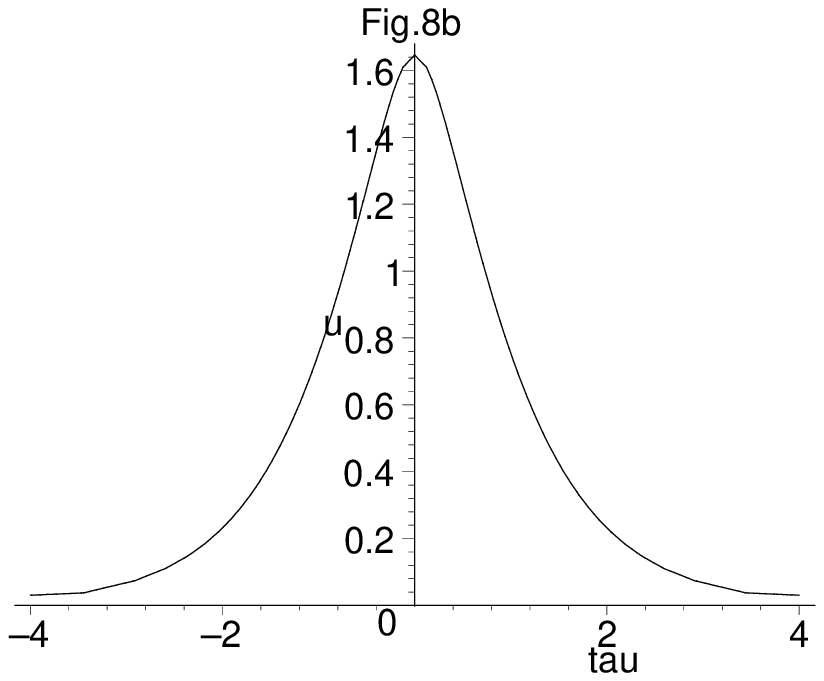} \caption{(a). The
typical structures for potential $V(u)$ with $c_1>0$, $c\geq0$,
$c_0^2<1,\ c^2>1$ and $\{c_0=\frac{1}{\sqrt{3}},\ c=2,\
c_1=\frac12\}$. (b). The corresponding solitary wave related to (a). }
\end{figure}

In the first three cases, $c_0$ is less than $1$ and the special
models \eqref{41} and \eqref{43} satisfy this condition. When this
condition is not satisfied, we have three other different
potential structures.

\em Case 4. \rm
\begin{equation}\label{case4}
0< c^2<1<c_0^2.
\end{equation}
The corresponding potential structure of the fourth case
\eqref{case4} is plotted in Fig. 9a. Different from the first and
third cases, from Fig. 9a we know the quasi-particle located the
$u=0$ hill may roll down the hill to both the left and right sides
and finally return back to the hill. So in this case, the right
moving solitary waves may have both bell and ring shapes and the
left moving solitary waves possess the same property. The right
moving bell shape solitary wave solution related to this case with
the same parameters as Fig. 9a is plotted in Fig. 9b. From Fig.
9a, we can know also that for the right moving solitary waves, the
bell shape solitary waves possess larger amplitudes and for the
left moving solitary waves, the ring shape solitary waves possess
larger amplitudes.

\input epsf
\begin {figure}
\epsfxsize=7cm\epsfysize=5cm\epsfbox{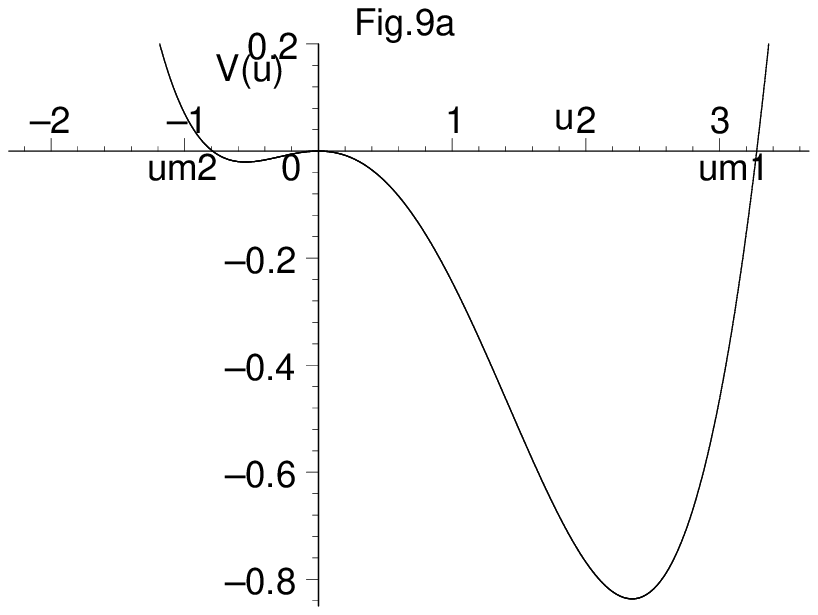}
\epsfxsize=7cm\epsfysize=5cm\epsfbox{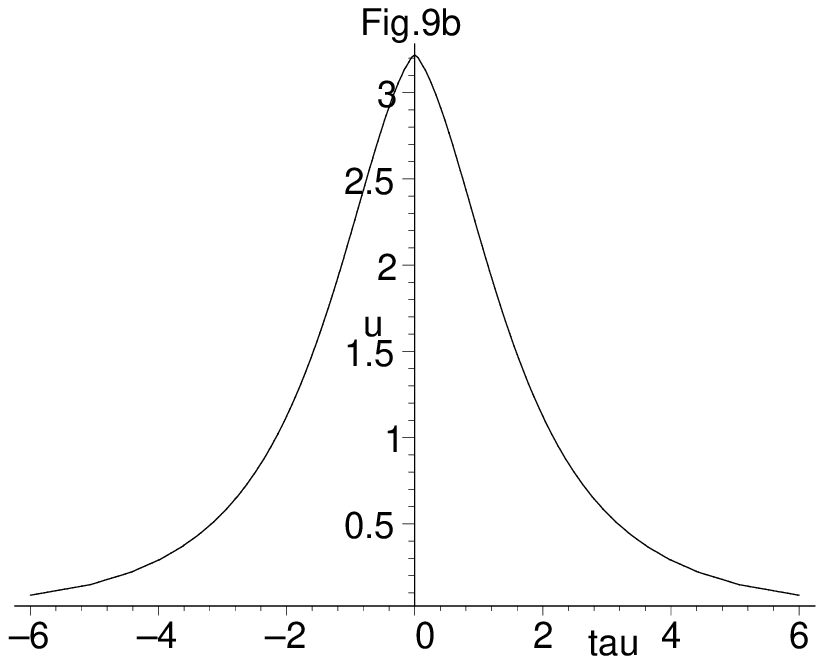} \caption{(a). The
typical structure for potential of the quasi-particle located in
with $c_1>0$, $c\geq0$, $c_0^2>1, c^2<1$ and $\{c_0=2,\ c=0.6,\
c_1=\frac12\}$. (b). The corresponding solitary wave related to (a). }
\end{figure}

\em Case 5. \rm
\begin{equation}\label{case5}
1< c^2<c_0^2.
\end{equation}
When $c^2$ is located in the range $(1,\ c_0)$, the corresponding
potential structure possesses the form as shown in Fig. 10. From Fig.
10 we know that $u=0$ is only a minimum of the potential. So similar to
the discussion of the case 2, we can conclude that there is no
zero boundary solitary wave for the GBQS with the condition
\eqref{case5}.

\input epsf
\begin {figure}
\epsfxsize=7cm\epsfysize=5cm\epsfbox{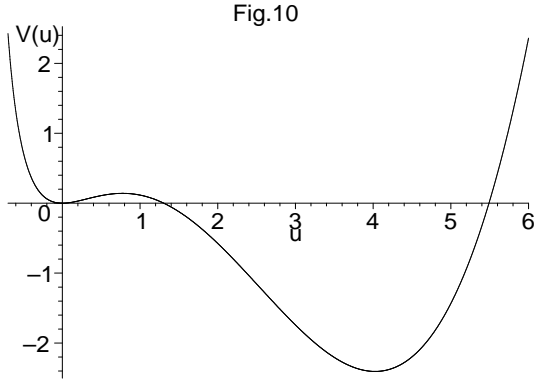} \caption{The
typical structures for potential $V(u)$ with $c_1>0$, $c\geq0$,
$1<c^2<c_0^2$ and $\{c_0=2,\ c=1.6,\ c_1=\frac12\}$.}
\end{figure}

\em Case 6. \rm
\begin{equation}\label{case6}
1< c_0^2<c^2.
\end{equation}
For the last case \eqref{case6}, the related potential structure
is shown in Fig. 11. From Fig. 11 we know that $u=0$ is really a
maximum of the potential, the quasi-particle may roll down the
$u=0$ hill in both sides, however, it can not be back! That means
the fast moving ($c^2>c_0$) solitary waves for the GBQS with the
condition $c_0^2>1$ are completely prohibited also.

\input epsf
\begin {figure}
\epsfxsize=7cm\epsfysize=5cm\epsfbox{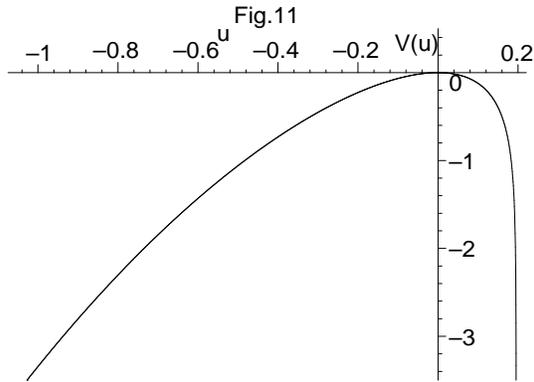} \caption{The
typical structure for potential $V(u)$ with $c_1>0$, $c\geq0$,
$1<c_0^2<c^2$ and $\{c_0=2,\ c=2.1,\ c_1=\frac12\}$.}
\end{figure}

Combining the critical cases, we summarize the conclusions for the
possible solitary waves of the GBQS \eqref{D} with zero boundary
conditions \eqref{boundary} and $c_1>0$ in the following table.

\newpage

\hskip.2in \bf Table 1. \rm Existent and prohibited regions of the
solitary waves.

\begin{tabular}{|c|c|c|c|c}
\hline\hline model & velocity & solitary wave & shape \\
\hline\hline
 $c_2<c_1$ & $c<-1 $ & yes & ring shape ($u<0$)\\ \hline
 $c_2<c_1$ & $-1\leq c\leq -|c_0|$   & no  &           \\ \hline
 $0<c_2<c_1$ & $ -|c_0|<c<0$   & yes & ring shape \\ \hline
 $c_2<0$ & $ -|c_0|<c<0$   & no &   \\ \hline
 $0<c_2<c_1$ & $c=0$ & yes &  both ring shape and bell shape \\ \hline
 $c_2<0$ & $c=0$ & no &   \\ \hline
 $0<c_2<c_1$ & $0<c<|c_0| $ & yes &  bell shape ($u>0$) \\ \hline
 $c_2<0 $ & $0<c<|c_0| $ & no &    \\ \hline
 $c_2<c_1$ & $|c_0|\leq c\leq 1 $ & no &   \\ \hline
 $c_2<c_1$ & $c>1$ & yes &  bell shape \\ \hline\hline
 $c_1<c_2$ & $c<-1 $ & no & \\ \hline
 $c_1<c_2$ & $c= -1$   & yes  &  algebraic, ring shape         \\ \hline
 $c_1<c_2$ & $ -1<c<1$   & yes & both ring shape and bell shape\\ \hline
 $c_1<c_2$ & $c=1$ & yes &  algebraic, bell shape     \\ \hline
 $c_1<c_2$ & $c>-1 $ & no & \\ \hline\hline
\end{tabular}

\section{Summary and discussions}

In summary, using the standard WTC's Painlev\'e PDE test, we prove
that the GBQS is non-Painlev\'e integrable except for the special
case for $c_1=0$, i.e., \eqref{45}. For the non-Painlev\'e
integrable GBQS with zero boundary conditions \eqref{boundary},
the truncated Painlev\'e expansion approach leads to only a
special $sech^2$ shape solitary wave solution with a special
velocity selection. To find all the possible travelling solitary
wave solutions of the GBQS with $c_1>0$ for all the possible model
parameter regions, we map the problem to find the possible motions
of a Newtonian classical quasi-particle moving in some possible
potentials. After considering all the possible motions of the
classical quasi-particle in the possible potentials related to the
maxima of the potential at $u=0$, all the possible travelling
solitary wave solutions related to zero boundary conditions are
found for all the possible velocities and the model parameter
ranges with $c_1>0$.

Similar to the integrable cases such as the KdV equation, the
solitary waves at some special ranges, the travelling solitary
waves with zero boundary conditions are completely prohibited. For
the KdV equation, all the left moving and static solitons with
zero boundary conditions are prohibited. For the GBQS with
$c_1>0$, there are three different cases. For the first case,
$0<c_2<c_1$, both the faster and slower moving solitary waves are
allowed while the solitary waves moving in the ``middle"
velocities $ \frac{c_2}{c_1}\equiv c_0^2\leq c^2 \leq 1$ are completely
prohibited. For the second case, $0<-c_2<c_1$, only the faster
moving solitary waves are allowed while all the slower moving
solitary waves $c^2<1$ are prohibited. For the third case
$0<c_1<c_2$, contrary to the second case, all the slower moving
solitary waves are allowed while all the faster moving $c^2>1$
solitary waves are completely prohibited. For the first two cases
the solitary waves at the critical velocities $c=\pm1$ are
prohibited while for the third case, there are algebraic solitary
waves at the critical velocities $c=\pm 1$. For the first two
types of GBQS, the right moving solitary waves possess bell shape
and the left moving solitary waves possess ring shape. For the
third type of GBQS, the right moving bell shape solitary wave
possess larger amplitudes than those of the right moving ring
shape solitary waves (with the same value of the velocity) while
for the left moving solitary waves, the ring shape solitary waves
possess larger amplitude.

\section*{Acknowledgements}
The work was supported by the National Outstanding Youth
Foundation of China (no. 19925522), the National
Natural Science Foundation of China (No. 90203001), the Research
Fund for the Doctoral Program of Higher Education of China (Grant.
No. 2000024832), the Natural Science Foundation of
Zhejiang Province of China and the ARC Large Grant A00103139 of
the University of New South Wales University.


\begin{thebibliography}{999}

\bibitem{li1} C. S. Gardner, J. M. Green, M. D. Kruskal, R. M. Miura,
  {sl Phys. Rev. Lett. }, 19:1095(1967)
\bibitem{wz} T. Y. Wu, and J. E. Zhang, on modelling nonlinear long
waves:in {\sl Mathematics is for solving problem: a volume in
honor of Julian Cole on his 70th birthday}, ed. L. P. Cook, V.
Roytburd and M. Tulin. SIMA, 233-249 (1996)
\bibitem{kaup} D. J. Kaup, {\sl Prog. Theor. Phys. } 54:396(1975)
\bibitem{kup} B. A. Kupershmidt, {\sl Comm. Math. Phys. } 99:51(1985)
\bibitem{ccl} R. Conte, M. Musette and A. Pickering, J. Phys. A:
Math. Gen. 27: 2831 (1994); M. Musette and R. Conte, J. Phys. A:
Math. Gen. 27: 3895 (1994); M. Musette, R. Conte and A. Pickering,
J. Phys. A: Math. Gen. 28: 179 (1995); A. Pickering, J. Math.
Phys. 37: 1894 (1996); C. L. Chen and S. Y. Lou, Chaos, Solitons
and Fractals, 16: 27 (2003).
\bibitem{zli} J. E. Zhang, and Y. S. Li, Bidirectional Solitons on
Water, in press
\bibitem{WTC} J. Weiss, M. Tabor, G. Carnevale, {\sl J. Math. Phys.}
             24: 522(1983).
\bibitem{Pain}S-y Lou, Phys. Rev. Lett., 80: 5027; S-y Lou ans J-j
Xu, J. Math. Phys. 39: 5364 (1998).
\bibitem{Lou}S-y Lou, Z. Naturforsch. 53a: 251 (1998).
\bibitem{chara} J. Weiss, Nato. Adv. Sci. I Ser. B 278: 225
(1992); S-y Lou, Phys. Lett. A176: 96 (1993); X-y Tang and H-c Hu,
Chin. Phys. Lett. 19: 1225 (2002).
\bibitem{instanton} R. Rajaraman, \em Solitons and Instantons, \rm
(North-Holland, Amsterdam, 1982), pp. 16-19.
\bibitem{sG} Y-j Zhu and S-y Lou, {\sl Commun. Theor. Phys.}, 23: 83
(1995); S-y Lou and G-j Ni, {\sl J. Math. Phys.} 30: 1614 (1989).
\bibitem{bkk} J. M. Dye and A. Parker, {\sl J. Math. Phys.} 43:
4921 (2002).
\end{thebibliography}
\end{document}